\documentclass[10pt,pra,twocolumn,showpacs,floatfix,superscriptaddress]{revtex4}
\usepackage{graphicx}
\usepackage{epstopdf}
\usepackage[colorlinks]{hyperref}
\hypersetup{
 colorlinks=true,
 citecolor=black,
 linkcolor=black,
 urlcolor=black}
\usepackage{amssymb}
\usepackage{times}
\usepackage{amsmath,mathrsfs}
\usepackage{amsthm}
\usepackage{amsfonts}%
\newcommand{\bra}[1]{\langle#1|}
\newcommand{\ket}[1]{|#1\rangle}
\renewcommand{\sec}[1]{\hyperref[sec:#1]{Sec.~\ref{sec:#1}}}
\newcommand{\eq}[1]{(\ref{eq:#1})}
\newcommand{\fig}[1]{\hyperref[fig:#1]{Fig.~\ref{fig:#1}}}

\begin{document}

\title{Device-Independent Quantum Cryptography for Continuous Variables}

\author{Kevin Marshall}
\email{marshall@physics.utoronto.ca}
\affiliation{Department of Physics, University of Toronto, Toronto, M5S 3G4, Canada}
\author{Christian Weedbrook}
\email{christian.weedbrook@gmail.com}
\affiliation{Department of Physics,
University of Toronto, Toronto, M5S 3G4, Canada}
\affiliation{QKD Corp., 60 St.~George St., Toronto, M5S 3G4, Canada
}

\date{\today}

\begin{abstract}

We present the first device-independent quantum cryptography protocol for continuous variables. Our scheme is based on the Gottesman-Kitaev-Preskill encoding scheme whereby a qubit is embedded in the infinite-dimensional space of a quantum harmonic oscillator.  The novel application of discrete-variable device-independent quantum key distribution to this encoding enables a continuous-variable analogue.  Since the security of this protocol is based on discrete-variables we inherit by default security against collective attacks and, under certain memoryless assumptions, coherent attacks.  We find that our protocol is valid over the same distances as its discrete-variable counterpart, except that we are able to take advantage of high efficiency commercially available detectors where, for the most part, only homodyne detection is required.  This offers the potential of removing the difficulty in closing the loopholes associated with Bell inequalities.

\end{abstract}

\maketitle

\section{Introduction}
\label{sec:intro}
Quantum key distribution (QKD) \cite{Gisin02,Scarani09} is a method by which two parties, Alice and Bob, may generate a shared secret key over an insecure quantum channel monitored by an eavesdropper, Eve.  Any QKD protocol relies on several assumptions, namely, any eavesdropper must obey the laws of quantum mechanics; Alice and Bob have the freedom to choose at least one of two measurement settings; and there is no classical information leaking from Alice or Bob's laboratories.  Most conventional QKD protocols further assume that Alice and Bob have near perfect control of their measurement devices as well as their state preparation.  Device-independant QKD \cite{Mayers04,Acin07,Ekert14} is a protocol that, remarkably, is free from making these additional assumptions; Alice and Bob need no knowledge of the inner workings of their devices nor even the dimension of the space their quantum states reside in.

In this paper, we use the novel approach of combining the encoding scheme of \cite{GKP} with the results of \cite{Mayers04,Acin07} to create a device-independent quantum cryptography protocol for continuous variables (CVs). CV quantum information offers higher efficiency detectors, cheap off-the-shelf components and the experimentally accessible Gaussian resources.  Furthermore, by encoding the CV space of a harmonic oscillator into a finite-dimensional code space we are able to take advantage of results which have previously only been applied to discrete-variable (DV) QKD.

The first proposals for continuous-variable QKD \cite{Weedbrook12} relied on `non-classical' states of light such as squeezed states \cite{Hillary00,Gottesman01}.  In fact, one of these protocols was proven unconditionally secure \cite{Gottesman01}.  As the field matured it was recognized that such non-classical states were not required and that the more experimentally available class of coherent states were sufficient \cite{Grosshans02,Weedbrook12}.

Device-independent QKD provides a way by which two parties may share a private key despite having no knowledge of the inner workings of their respective devices.  Conversely, in conventional QKD protocols it is regularly assumed that both parties have a high degree of control over both state preparation as well as measurement. Although, recently relaxing the condition of trusting the measurement device was achieved \cite{Lo12,Braunstein12}. The security in this device-independent approach comes instead from the fact that the two parties are able to violate a Bell inequality \cite{Brunner13}, which can remarkably be used to put a bound on the amount of information that a potential eavesdropper could, in principle, obtain.

Here we introduce a CV version of device-independent QKD. Our protocol goes as follows. Alice first generates a Bell state which has been created using an encoding based on the Gottesman-Kitaev-Preskill scheme where a qubit is encoded into an infinite dimensional space of a harmonic oscillator. After this the protocol continues in a similar fashion where she keeps one encoded qubit and sends the other qubit to Bob over an insecure quantum channel.  Hence, the results of DV device-independent QKD can then be applied to the system yielding the first implementation of device-independent QKD for CVs.  

It is known in the field of CV quantum information that all Gaussian resources are insufficient for violating a Bell inequality \cite{Bell87,Paternostro09}.  This means that one should already expect non-Gaussian states or measurements as being a requirement \cite{Wenger03,Garcia05,Cavalcanti07,Brask12,Qian14}, despite the fact that they are typically more difficult to produce in a lab.  This highlights the challenges faced when attempting to create a CV version of device-independent QKD because most current CV-QKD protocols use Gaussian states. Fortunately, if we use for example, a single mode of the electromagnetic field as our harmonic oscillator, we are able to use CV resources, including high efficiency detectors and off-the-shelf components.  The major drawback of DV device-independent QKD is that in order to close the detector loophole one needs high efficiency detectors \cite{Rowe01}.  The detector loophole issue is often overcome by CV quantum information where we can take advantage of such high detection efficiencies~\cite{Wenger03,Garcia05,Cavalcanti07,Brask12,Qian14,Reid13}.

This paper is structured as follows. In \sec{background} we discuss separately the necessary encoding scheme as well as the results of DV device-independent QKD.  In \sec{cvdiqkd} we relate these concepts to CV quantum information and discuss formally the kind of measurements that are necessary.  Following this we investigate the resources required in order to implement the protocol in \sec{resources}.  Since we are only capable of making approximations of the desired encoding in the real world, we consider the effects of such approximations on the encoding and resulting key rate in \sec{finitesq}.  Finally, \sec{conclusions} presents some discussions and concluding comments as well as some interesting open questions.
\section{Background}
\label{sec:background}
The premise of this paper is to propose a method of implementing device-independent QKD with CV states. This is accomplished by embedding a two-level Hilbert space into the full infinite-dimensional space and then using results from DV-QKD.  Here we discuss the encoding scheme proposed by Gottesman, Kitaev, and Preskill (GKP) in \cite{GKP} as well as the DV version of device-independent QKD.
\subsection{GKP encoding}
\label{sec:encoding}
The GKP encoding \cite{GKP} provides a method to encode a qubit in the infinite-dimensional space of an oscillator in such a way that one can protect against arbitrary, but small, shifts in the canonical variables $q$ and $p$ as well as carrying out fault-tolerant universal quantum computation on the encoded space \cite{GKP,Menicucci14}.  The stabilizer generators of a two-dimensional Hilbert space in an infinite-dimensional Hilbert space with canonical variables $q,p$ are given by~ \cite{Gottesman01}
\begin{align}
S_q=\exp(2iq\sqrt\pi), \hspace{1mm} S_p=\exp(-2ip\sqrt\pi).  
\end{align}
The stabilizers are simply shift operators for $q,p$, and if the eigenvalues are $S_q=S_p=1$ then the allowed values of $q$ and $p$ are integer multiples of $\sqrt \pi$.  Since the codewords are invariant under shifts by integer multiples of $2\sqrt \pi$ we can define a basis for the encoded qubit as 
\begin{align}
\ket{\bar j_L}\propto \sum_{s\in\mathbb Z} \ket{(2s+j)\sqrt\pi}_q, 
\end{align}
for $j=0,1$, and where the subscript $q$ indicates the $q$ (`position')-basis.  These states can be approximated optically using Schr\"odinger cat states \cite{Vasconcelos10}, or by a variety of other methods \cite{Travaglione02,Pirandola04,Pirandola05,Pirandola06}.  Encoded Pauli gates are defined as $\bar Z\equiv \exp(iq\sqrt\pi)$ and $\bar X\equiv \exp(-ip\sqrt\pi)$; since these operators commute with the stabilizers they also preserve the code subspace.

The set of Clifford operations on the encoded subspace correspond to symplectic (or Gaussian) transformations on the CV space of the oscillator; these operations can be implemented in a fault tolerant way \cite{GKP}.  To achieve universal quantum computation we must be able to implement a non-Clifford gate on the encoded subspace \cite{Lloyd99}, for example, the addition of a $\pi/8$-gate (T-gate) to the Clifford group will make for a universal set of gates.  The T-gate can be implemented with a non-symplectic transformation on the oscillator; this is more experimentally difficult than symplectic transformations and requires a non-Gaussian resource such as photon counting.  The physical resources required to implement these gates are discussed in more detail in \sec{resources}.
\subsection{Device-independent quantum key distribution}
\label{sec:diqkd}
The DV-QKD protocol \cite{Mayers04,Acin07} begins with Alice and Bob sharing a quantum channel that emits pairs of entangled particles.  To consider the worse case scenario, we allow Eve full control over the source \cite{Lo99} which, if she is honest, emits the state $\ket{\psi_{AB}}=1/\sqrt 2(\ket{00}+\ket{11})$.  But in general she is free to create any arbitrary state $\rho_{ABE}$ which may be entangled between not only Alice and Bob but herself as well.  To generate a secret key, Alice chooses a basis to measure in from $\{A_0,A_1,A_2\}$ while Bob chooses a basis from $\{B_1,B_2\}$ and they get outcomes of $a_i,b_j\in \{+1,-1\}$, respectively~\cite{Pironio09}.  After all measurements are performed, if Alice had chosen measurement $A_0$ and Bob chosen measurement $B_1$ they extract a single bit of raw key corresponding to their measurement outcome.  Instead, if they had measurement settings corresponding to $\{A_0,B_2\}$, their outcomes are completely uncorrelated and so this case is discarded.  For all other measurement settings Alice and Bob use their results to violate the CHSH inequality \cite{Brunner13} 
\begin{align}
\mathcal S&=\langle a_1b_1 \rangle +\langle a_1 b_2 \rangle + \langle a_2b_1 \rangle - \langle a_2b_2 \rangle\leq 2.
\end{align}
The CHSH inequality puts a bound on the values of $\mathcal S$ consistent with local hidden-variable theories in accordance with Bell's theorem \cite{Bell,CHSH}.  Violation of this inequality by quantum mechanics arises due to the fact that entanglement can provide nonlocal correlations that cannot be produced by shared randomness.  If Alice and Bob share a nonlocal correlation then, regardless of how this correlation came to exist, Eve cannot have full knowledge of the correlation or else she would be in possession of a local variable capable of reproducing the correlations \cite{Acin07}.

A set of measurements which give the desired behaviour in the above protocol and which maximize the violation of the CHSH inequality are given by \cite{Pironio09}
\begin{align}
\begin{aligned}
\label{eq:bases}
A_0=B_1&= Z, & A_1&= 1/\sqrt 2 ( Z +  X),\\
B_2&= X, & A_2&=1/\sqrt 2( Z - X).
\end{aligned}
\end{align}
For the moment, $Z$ and $X$ in the above expression \eq{bases} have no relation to the encoded Pauli gates $\bar Z$ and $\bar X$; although we will make this connection in \sec{cvdiqkd}. The main result shown by Ac\'in et al. \cite{Acin07} is that the Holevo quantity between Eve and Bob, after Alice and Bob have symmetrized their marginals, is bounded as
\begin{align}
\chi(B_1{:}E)&\leq h\left(\frac{1+\sqrt{(\mathcal S/2)^2-1}}{2}\right),
\end{align}
where $\chi(B_1{:} E)=S(\rho_E)-\frac{1}{2}\sum_{b_1=\pm 1}S(\rho_{E|b1})$ is the Holevo quantity and $h=-p\log_2p-(1-p)\log_2(1-p)$ is the binary entropy.  This provides a method which Alice and Bob can use to keep Eve honest and bound her knowledge using only their violation of the CHSH inequality.
\section{Continuous-Variable Device Independence Protocol}
\label{sec:cvdiqkd}
The CV version of device-independent QKD begins with Alice creating an encoded Bell state.  This encoding is based on the GKP encoding as given in \sec{encoding}.  Once this Bell state is created she keeps one qubit for herself and sends the other entangled qubit to Bob over an insecure and lossy quantum channel. Apart form this initial encoding, the protocol follows the same steps as in typical DV-QKD protocols \cite{Mayers04,Acin07}.

A set of measurements which maximize the violation of the CHSH inequality, for the encoded state $\ket{\Phi^+}=1/\sqrt 2(\ket{\bar 0\bar 0}+\ket{\bar 1 \bar1})$, consist of measurements $A_1,A_2,B_1,B_2$, as defined in \sec{diqkd}, which act on the encoded subspace. We can destructively measure the observables $\bar Z$ and $\bar X$ by performing a suitable homodyne measurement of the $\hat q$ or $\hat p$ quadrature, respectively. By measuring the $\hat q$ quadrature we expect that the only outcomes possible will be integer multiples of $\sqrt\pi$; even multiples corresponding to a $\ket{\bar 0}$ state and odd multiples corresponding to $\ket{\bar 1}$.  Imperfections in the measurement and the encoded state will result in other measured values, but we can apply classical error correction and adjust the value to the nearest $k\sqrt\pi$ for an integer $k$.  The outcome of the measurement $\bar Z$ is then given by $(-1)^k$.  We can measure the other three observables by first applying a change of basis gate which takes us to the $\bar Z$ basis, and in this way we need only consider homodyne measurements of the $\hat q$ quadrature.  

We assume in this section that we are able to implement Clifford gates as well as $\pi/8$-gates on our encoded space and also that we can carry out homodyne measurements on the CV space; the resources required to do this are discussed in \sec{resources}.  From here onwards we drop the over-bar notation to denote encoded operations; all gates are to be understood as acting on the encoded space while symplectic transformations are understood to be in relation to the oscillator. 

It is readily seen that we can measure in the $X$ basis by using the change of basis gate $H$, and one can easily verify that $ H^\dagger  X  H= Z$.  Since the Hadamard gate $H$ is in the Clifford group we can implement an encoded $H$ by carrying out symplectic transformations on the full CV space.  Unfortunately, it is not possible to change from the $A_1$ or $A_2$ basis to the $Z$ basis by using only Clifford gates, which means that we will need to go beyond symplectic transformations in the CV space.  This can be readily seen by recognizing that $A_1= H$, suppose there existed a Clifford gate $C$ such that $C^\dagger  H C= Z$.  This would imply that $C Z C^\dagger= H$ and so $C$ is not a Clifford gate by definition.

The required change of basis gates can be calculated as: $\mathcal I B_1 \mathcal I = Z,~H^\dagger B_2 H=Z,~\alpha^\dagger A_1 \alpha=Z$, and $\beta^\dagger A_2\beta =Z$, where $\alpha=PHTHP$, $\beta=ZPHTHP$ and $T$ is an encoded $\pi/8$-gate.  It is important to note that while the latter two gates are not Clifford gates, they can be decomposed exactly as a composition of Clifford gates with only one non-Clifford $T$-gate.  Furthermore, it is possible to shift the problem of implementing a $T$-gate to a state preparation problem, and since preparation can be done `offline' we require only Gaussian operations and one auxiliary state to carry out our CV device-independent QKD.

\section{Required resources}
\label{sec:resources}
In order for Alice and Bob to implement the necessary measurements they must be able to perform gates on the encoded states as well as homodyne detection on one quadrature.  The necessary set of gates include $ H,  P,  T$ (no need for $Z$, since $ Z= P^2$). The first two gates correspond to Clifford operations while the last one is a non-Clifford gate.  The set of Clifford gates on the encoded states correspond to symplectic transformations on the CV space, given as \cite{GKP}: $H: (q,p)\rightarrow (p,-q)$, $P: (q,p)\rightarrow (q,p-q)$, and $C_{NOT}: (q_1,p_1,q_2,p_2)\rightarrow (q_1,p_1-p_2,q_1+q_2,p_2)$. The encoded $C_{NOT}$ gate is used not by Alice or Bob but in the preparation of the encoded Bell state by Eve.

In order to implement an encoded $\pi/8$-gate we need a non-symplectic transformation which requires a non-Gaussian resource.  The addition of photon counting to Gaussian resources is sufficient to carry out non-symplectic transformations.  In particular, one is able to create either a $\pi/8$ state or a cubic phase state which can then be used to implement a $T$-gate on the encoded space \cite{GKP,Gu09}.  Fortunately, one can generate these states offline and use them as required throughout the protocol, effectively shifting the issues of non-Gaussian operations to state preparation.  In this way, one needs only to have a supply of non-Gaussian states and be capable of performing symplectic transformations (including homodyne detection) in order to implement the QKD protocol.  In the case of an optical mode, the set of symplectic transformations can be achieved with linear optics (phase shifters and beam splitters) and squeezing operations (non-linear crystals).

Fortunately, Alice and Bob do not need to choose a measurement basis, which is used to check for a CHSH violation, very often; the probability to choose between the possible options need not be uniform, although this would work as well.  If we suppose that Alice and Bob share $N$ quantum states, it is enough to use $\sim \sqrt N$ pairs to check for a CHSH violation, so long as the measurements are causally independent \cite{Masanes11}.  This condition would be satisfied for memoryless devices, or devices which could have internal memory reliably cleared after every run.  One protocol \cite{Masanes11}, also provides security against coherent attacks, which is the most general form attack.  Since we have chosen the measurement basis corresponding to generating a key as $Z$, this means that in the limit of large $N$ almost all of the time we need only perform Gaussian operations.  Hence, we need only perform non-Clifford operations a small fraction of the time, in order to estimate the CHSH violation and thus keep the eavesdropper honest.  Many other such DV device-independent QKD protocols exist and offer different key rates with different underlying assumptions \cite{Barret05,Lim13,Vazirani12}, but typically one still requires the ability to make measurements in a set of four bases which violates the CHSH inequality.
\section{Gaussian finite-squeezing effects}
\label{sec:finitesq}
In practice, the encoded GKP states will not consist of delta peaks at $\sqrt\pi$ intervals, but instead the peaks will have some finite width and they will be modulated by a larger envelope to ensure the state is of finite energy.  One way to produce an ideal GKP state is to prepare a momentum eigenstate $\ket{p=0}$, and then measure the value of $q~ \text{(mod 2}\sqrt\pi)$.  Since the position is completely undetermined for a momentum eigenstate all values of $q$ are equally likely, and this measurement will project out a state that differs from a $ Z$ eigenstate by a shift of $q$ which can then be corrected.

If instead of an unphysical momentum eigenstate, which corresponds to infinite squeezing, we can consider a finitely-squeezed state given by $\psi_{sq}(p)=\pi^{-1/4}\kappa^{-1/2}\exp\left(-\frac{1}{2}p^2/\kappa^2\right)$ \cite{Braunstein05}, where $\kappa=e^{-r}$ for squeezing parameter $r\in\left[0,\infty\right)$.  In the position representation this state is given by $\psi_{sq}(q)=\pi^{-1/4}\kappa^{1/2}\exp(-\frac{1}{2}q^2\kappa^2)$.
An ideal homodyne measurement of $q$ is a projection-valued measure (PVM) with projectors corresponding to position eigenstates $P_x=\ket x\bra x$, or infinitely squeezed states in position.  If we allow the homodyne measurement to have a Gaussian acceptance of width $\Delta$, we replace the PVM with a positive-operator valued measure (POVM) which consists of an ideal homodyne measurement convolved with a Gaussian window.  This leads to POVM elements given by
\begin{align}
\label{eq:povm}
\Pi_x&=(2\pi\Delta^2)^{-1/2}\int_{-\infty}^\infty dy e^{-\frac{1}{2}(x-y)^2/\Delta^2}\ket y\bra y.
\end{align}
An ideal measurement of $q~ \text{(mod 2}\sqrt\pi)$ is described by the PVM with elements $P'_x=\sum_{s=-\infty}^\infty P_{x-2s\sqrt\pi}$ for $x\in[0,2\sqrt\pi)$, and if we let $P_x\rightarrow \Pi_x$ we obtain the result for a homodyne detector with a Gaussian acceptance.  Without loss of generality suppose we obtain a result corresponding to $\Pi_0$, then the state will be transformed to $\psi_{\bar 0}(q)\propto \sum_{s=\mathbb Z} \exp(-\frac{1}{2}q^2\kappa^2) \psi'_{sq}(q+2s\sqrt\pi)$,
where $\psi'_{sq}$ is a squeezed vacuum state in position with width $\Delta$. If we obtain a result other than $\Pi_0$ we can simply apply a shift to correct the state. This is of the same type of approximate codeword proposed in the GKP paper \cite{GKP}.  Notice that our initial squeezing determines the size of the overall envelope, width $\kappa^{-1}$, while the precision of our homodyne measurement determines the width of the individual peaks.

If we further approximate by replacing $\exp(-\frac{1}{2}q^2\kappa^2)\rightarrow \exp(-\frac{1}{2}(2s\sqrt\pi)^2\kappa^2)$ in the summation above, which corresponds to scaling each peak by a constant factor, we find
\begin{align}
|\psi_{\bar 0}(q)|^2&=\frac{2\kappa}{\Delta\sqrt{\pi}}\sum_{s=-\infty}^\infty e^{-4\pi\kappa^2 s^2}e^{-(q-2s\sqrt\pi)^2/\Delta^2}.
\end{align}
We can correct for shifts in the position which are less than $\sqrt\pi/2$, and thus bound the error by adding up the contribution from all of the tails further than $\sqrt\pi/2$ from their respective peak.  Assuming that $\kappa\sqrt\pi\ll 1$ the probability of error is bounded as $P_e<2\Delta^2/(\kappa\pi)\exp{(-\frac{1}{4}\pi/\Delta^2)}$ \cite{Gottesman01}.

\begin{figure}[htp]
\centering
\includegraphics[scale=0.84]{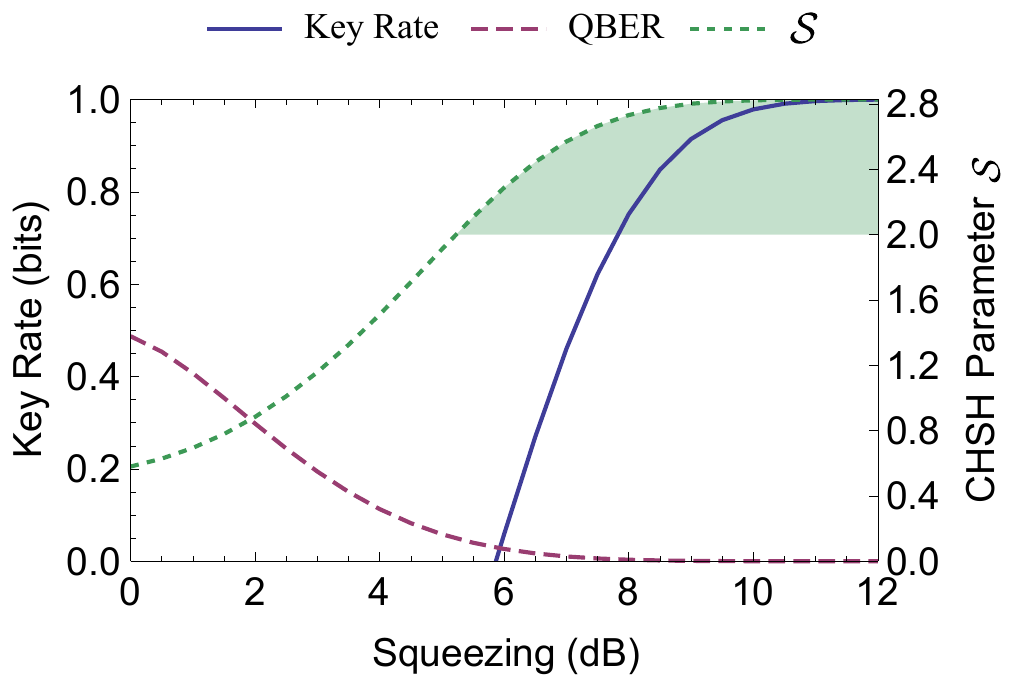}
\caption{(color online) The extractable secret-key rate is plotted as a function of the squeezing for the symmetric case $\Delta=\kappa$, where $\Delta$ is the width of the individual peaks and $\kappa^{-1}$ is the width of the Gaussian envelope in the GKP encoding.  Note that currently the maximal amount of single-mode squeezing achieved is $12.7$ dB \cite{Eberle10,Mehmet11}.  The shaded region indicates a violation of the CHSH inequality.}
\label{fig:keyrate}
\end{figure}

The errors from incorrectly identifying an encoded state will determine the amount by which one is able to violate the CHSH inequality.  Consider one term in the CHSH quantity $\mathcal S$. The correlator is defined as $\langle a_ib_j\rangle=P(a=b|ij)-P(a\neq b|ij)$ for outcomes $a,b$ and measurement choices $i,j$.  If we assume that our gates are perfect then all errors will come from incorrectly identifying an encoded state.  We can calculate the value of $\mathcal S$ after error correction by computing the expectation values of the various measurements.  This value is plotted in \fig{keyrate}, and it can be seen that we start to violate the CHSH inequality for parameters $\Delta=\kappa$ corresponding to squeezing greater than 5 dB. This shows that the value of the CHSH quantity is scaled according to the error rate, assuming perfect gates.

The quantum bit error rate (QBER) \cite{Scarani09} is defined as $Q=P(a\neq b|01)=2P_e(1-P_e)$ since we are only extracting a key for the cases where Alice does measurement $A_0$ and Bob does measurement $B_1$.  This corresponds to either Alice or Bob incorrectly identifying the state while the other party does not make an error.  The secret-key rate $r$, under collective attacks, with one-way classical post-processing from Bob to Alice, is lower bounded by the Devetak-Winter rate \cite{Acin07,Devetak05}
\begin{align}
r\geq r_{DW}=I(A_0{:} B_1)-\chi(B_1{:} E),
\end{align}
where $I(A_0{:}B_1)=1-h(Q)$ is the mutual information between Alice and Bob ($h$ being the binary entropy), and $\chi(B_1{:} E)$ is the Holevo quantity. 

In \fig{keyrate}, we plot both the QBER and the key rate $r$.  Notice that the extractable key rate remains zero even for values of $\mathcal S$ slightly larger than two. The key rate grows rapidly for squeezing beyond 6 dB, for example, a squeezing of 10 dB yields a key rate of $\approx 98\%$.  Note that the well known critical QBER of $11\%$ for BB84 \cite{Shor00} as well as $7.1\%$ for DV device-independent QKD \cite{Acin07} are higher than the $\approx 3.5\%$ critical QBER for this proposal.  This is due to the fact that one requires a suitable enough approximation to a GKP encoded state in order to have a high enough violation of the CHSH inequality, and by doing so one immediately achieves a corresponding low probability of error $P_e$.  Typically one desires a high critical QBER as it generally tolerates more imperfections in the protocol.  However, in this case the difficulty arises from the need to violate the CHSH inequality and if one is able to do so then one already obtains a small QBER.  Intuitively, as the width of the individual peaks $\Delta$ in the encoded state become larger, and equivalently the QBER, the overall state resembles a Gaussian state and thus cannot violate the CHSH inequality.
\section{Discussion of loss, comparison to discrete-variables and conclusion}
\label{sec:conclusions}
By harnessing the results of discrete-variable QKD, with a qubit encoding in a harmonic oscillator, we provided the first device-independent QKD protocol for continuous variables.  This protocol derived its security from the ability to violate a Bell inequality and, remarkably, does not require Alice or Bob to know the inner workings of their devices.  We showed how both the CHSH violation and the resulting extractable key-rate depended on the quality of the approximate codewords. We also showed that, in terms non-Gaussian resources, we required a modest one $T$-gate for each of $\sim \sqrt N$ of $N$ total Bell pairs.  Thankfully, from an experimental point of view, what this means is that only Gaussian operations (e.g., homodyne detection) are needed most of the time.

It should be noted that our encoding scheme is experimentally challenging.  However, it is still practical, with many proposals already existing \cite{Vasconcelos10,Travaglione02,Pirandola04,Pirandola05,Pirandola06}.  It is hoped that our paper will further motivate experimental advances using such encodings.  Given the technological challenges, distances in our scheme will be limited (although not fundamentally).  However, it should be noted that such limitations are also faced by the discrete-variable version of device-independent QKD, which is currently limited to a few kilometers.  This is because it requires a detection efficiency of approximately $95\%$ to achieve a key rate on the order of $10^{-10}$ per pulse \cite{Marcos11}.  

Interestingly, one can also consider the distances over which our continuous-variable protocol will perform well. We do this by calculating the Wigner function of an approximate encoded state and then send it through an amplitude damping channel. We numerically find that, for example, at 2.3km, with 0.2dB/km loss, we get a key rate of 0.35 bits/state. This is comparable to discrete-variable device-independent QKD where such schemes are limited to only a few kilometers \cite{Marcos11}.  Furthermore, we note that the distance of our protocol can also be improved by using distillers as was shown for discrete-variable states \cite{Gisin10} or by the application of heralded amplifiers~\cite{Lutkenhaus11}.



It is an interesting open question whether one can devise a device-independent continuous-variable QKD protocol with more readily accessible states.  One possible avenue to explore is lifting the requirement of a CHSH inequality violation by considering a protocol where only one party trusts their device.  This one-sided device-independent QKD requires one to violate only an EPR-steering inequality \cite{Reid13}, which amounts to Alice and Bob checking that they have entanglement and ruling out local hidden state models \cite{Branciard12}.  

C.W. acknowledges support from NSERC.  We are grateful to Hoi-Kwong Lo and Norbert L\"utkenhaus for fruitful discussions.
\twocolumngrid
\appendix*
\section{}
Here we justify the use of several results in the paper.  In \sec{finitesq} we identified the approximate encoded state $\psi_{\bar 0}(q)$ that would result when a finitely squeezed state $\psi_{sq}(p)=\pi^{-1/4}\kappa^{-1/2}\exp\left(-\frac{1}{2}p^2/\kappa^2\right)$ was measured with a POVM Eq.~\eq{povm} subject to $q$ mod $2\sqrt\pi$.  Supposing that we find an outcome corresponding to $\Pi_0$ we have
\begin{align}
\ket{\psi_{\bar 0}}&\propto\sum_{s=-\infty}^\infty(2\pi\Delta^2)^{-1/2} \int_{-\infty}^\infty dy e^{-\frac{1}{2}(2s\sqrt\pi+q)^2/\Delta^2}\nonumber\\
&~\times \pi^{-1/4}\kappa^{1/2}e^{-\frac{1}{2}q^2\kappa^2}\delta(q-y),\nonumber\\
&\propto \sum_{s=-\infty}^\infty e^{-\frac{1}{2}q^2\kappa^2}  e^{-\frac{1}{2}(2s\sqrt\pi+q)^2/\Delta^2},\nonumber\\
&\propto \sum_{s=-\infty}^\infty e^{-\frac{1}{2}q^2\kappa^2} \psi'_{sq}(q+2s\sqrt\pi).
\end{align}
Thus we recover the fact that this measurement projects the state onto a superposition of squeezed states in $\hat q$ with a spacing of $2\sqrt\pi$ and width of $\Delta$ weighted by an overall Gaussian envelope of width $\kappa^{-1}$.

In \sec{finitesq} we stated that the value of $\mathcal S$, with error correction, can be calculated by finding the appropriate expectation values.  Consider the box function
\begin{align}
\Pi(x)= \left\{
     \begin{array}{lr}
       1 & \quad \text{if\quad} |x| \leq 1/2\\
       0 & \quad \text{if\quad} |x| > 1/2
     \end{array}
   \right. .
\end{align}
To calculate the value of Tr$(\ket m\bra n Z)$, with $m,n \in \left\lbrace 0,1\right\rbrace$, for an element of the density matrix, we simply perform the integral
\begin{align}
\int_{-\infty}^\infty \psi_m(q)Z(q)\psi_n(q)dq,
\end{align}
where $Z(q)=2\left\lbrace\sum_{s\in Z} \Pi[(q-2s\sqrt\pi)/\sqrt\pi]-1/2\right\rbrace$.  Calculating a trace involving $X$ is similar to  $q$ being replaced by $p$ and using the corresponding Fourier transforms of the wavefunctions, while the other two necessary measurements 
\begin{align}
\alpha&=\left( \begin{array}{ccc}
-i\cos\frac{\pi}{8} & -i\sin\frac{\pi}{8}\\
-i\sin\frac{\pi}{8} & \phantom{-}i\cos\frac{\pi}{8}\\
\end{array} \right),\\
\beta&=\left( \begin{array}{ccc}
-i\cos\frac{\pi}{8} & -i\sin\frac{\pi}{8}\\
\phantom{-}i\sin\frac{\pi}{8} & -i\cos\frac{\pi}{8}\\
\end{array} \right),
\end{align}
can be decomposed into a sum of $Z, X$ and calculated using the linearity of the trace function.
\newpage
%

\end{document}